\documentclass[useAMS,usenatbib]{mn2e}
\usepackage{graphicx,natbib,color,multirow,amsmath,url,soul}
\usepackage{epsfig}
\usepackage{float}
\usepackage{deluxetable}

\newcommand\ion[2]{[#1$\;${\scshape{#2}}]}      
\newcommand\pfeatures{$p_{\rm{features~or~disk}}$}
\newcommand\plusminus[2]{\genfrac{}{}{0pt}{}{#1}{#2}}
\newcommand\pnotedgeon{$p_{\rm{not~edge-on}}$}
\newcommand\pbar{$p_{\rm{bar}}$}

\newcommand\gztwo{Galaxy~Zoo~2}
\newcommand\mbh{$M_{\rm{BH}}$}
\newcommand\db{$d_{\rm{B-NB}}$}
\newcommand\fb{$f_{\rm{B>NB}}$}

\begin{document}

\title[Galaxy~Zoo: bars and AGN]{Galaxy~Zoo: the effect of bar-driven fueling on the presence of an active galactic nucleus in disc galaxies}
\author[Galloway et~al.]{\parbox[t]{16cm}{Melanie A. Galloway$^1$, Kyle W. Willett$^1$, Lucy F. Fortson$^1$, Carolin N. Cardamone$^{2}$, Kevin Schawinski$^3$, Edmond Cheung$^{4,5}$, Chris J. Lintott$^6$, Karen L. Masters$^{7,8}$, Thomas Melvin$^7$, Brooke D. Simmons$^6$
\vspace{0.1in} }\\
$^{1}$School of Physics and Astronomy, University of Minnesota, 116 Church St. SE, Minneapolis, MN 55455, USA\\
$^{2}$Wheelock College, Department of Science, Wheelock College, Boston, MA 02215, USA\\
$^{3}$Institute for Astronomy, Department of Physics, ETH Z\"urich, Wolfgang-Pauli-Strasse 16, CH-8093, Z\"urich, Switzerland\\
$^{4}$Department of Astronomy and Astrophysics, 1156 High Street, University of California, Santa Cruz, CA 95064, USA\\
$^{5}$Kavli IPMU (WPI), The University of Tokyo, Kashiwa, Chiba 277-8583, Japan\\
$^{6}$Oxford Astrophysics, Denys Wilkinson Building, Keble Road, Oxford OX1 3RH, UK\\
$^{7}$Institute of Cosmology \& Gravitation, University of Portsmouth, Dennis Sciama Building, Portsmouth PO1 3FX, UK\\
$^{8}$SEPnet --- \url{http://www.sepnet.ac.uk} \\
   }
\maketitle

\begin{abstract}

	We study the influence of the presence of a strong bar in disc galaxies which host an active galactic nucleus (AGN). Using data from the Sloan Digital Sky Survey and morphological classifications from the \gztwo~project, we create a volume-limited sample of 19,756 disc galaxies at $0.01<z<0.05$ which have been visually examined for the presence of a bar. Within this sample, AGN host galaxies have a higher overall percentage of bars (51.8\%) than inactive galaxies exhibiting central star formation (37.1\%). This difference is primarily due to known effects; that the presence of both AGN and galactic bars is strongly correlated with both the stellar mass and integrated colour of the host galaxy. We control for this effect by examining the \emph{difference} in AGN fraction between barred and unbarred galaxies in fixed bins of mass and colour. Once this effect is accounted for, there remains a small but statistically significant increase that represents 16\% of the average barred AGN fraction. Using the $L_{\rm{[O~III]}}$/\mbh~ratio as a measure of AGN strength, we show that barred AGN do not exhibit stronger accretion than unbarred AGN at a fixed mass and colour. The data are consistent with a model in which bar-driven fueling does contribute to the probability of an actively growing black hole, but in which other dynamical mechanisms must contribute to the direct AGN fueling via smaller, non-axisymmetric perturbations. 

\end{abstract}

\section{Introduction}
\label{sec:Intro}

Supermassive black holes exist at the centres of most (if not all) massive galaxies \citep{KR95,richstone98,KG01,ghez08}. The evolution of the black hole is closely tied to that of the host galaxy; hence, understanding the conditions that drive black hole growth is key for a complete picture of galactic evolution. While most black holes are not actively growing, a small fraction are observed to accrete matter and cause the surrounding material to emit powerful pan-chromatic radiation. The central region of a galaxy which encompasses these ``active'' black holes, along with the surrounding accretion disk and ionized gas clouds, is an active galactic nucleus (AGN). Since the bolometric luminosity of the AGN can be comparable to (or greater than) the integrated stellar luminosity (as high as $L\sim10^{47}~\rm{erg}~\rm{s}^{-1}$) the black holes have an important effect on the host galaxy, controlling the amount of star formation via AGN feedback, as well as contributing toward the net reionization of the intergalactic medium \citep{HeckmanReview}. Understanding the fueling mechanism(s) for AGN is thus critical for studying galaxies, both in the nearby Universe and at higher redshifts. 

The precise physics that govern the relationship between AGN and their host galaxies is an area of intense study. This includes the AGN fueling mechanism --- while there is strong evidence that there is sufficient gas in the ISM to keep the accretion disc supplied with enough material to radiate at typical bolometric AGN luminosities \citep{Shlos89,Shlos90}, the dynamical mechanisms that drive the gas within the black hole's sphere of influence are difficult to observe directly, especially at extragalactic distances. In order to initiate (or continue) AGN activity, gas must lose enough angular momentum in a short timeframe to reduce its orbit from scales of kiloparsecs down to parsecs. \citet{Shlos89} analytically showed that while gas can lose angular momentum due to turbulent viscous processes, these are too slow to be the only mechanism involved. Later N-body simulations have shown viscous torques on the gas are negligible and do not directly initiate inflows \citep{Bournaud05}, further arguing for an additional method of radial gas transport. 

One possibility is that the presence of a large-scale bar may supplement viscous torques and further drive AGN fueling. Bars efficiently transport angular momentum within the disc \citep{Athanassoula03,KK04}, and are ubiquitous features in disc galaxies in the local Universe \citep{Eskridge2000,Laurikainen2004,Menendez2007,Masters11,Cheung2013}. Simulations \citep{Athanassoula92,FandB,AnnThakur05} show that stellar bars, whose lengths are on the order of kiloparsecs, do drive gas into the circumnuclear region (scales of 100~pc) of galaxies; observational studies have also shown an increase in the amount of central star formation for barred galaxies \citep{Ellison11}. This combination of simple analytical models, simulations, and observations clearly points toward galactic bars preferentially driving gas to the centres of their galaxies. It is still an open question, though, whether this gas is ultimately driven to the central $1-10$~pc scales, which theoretical models suggest are required for accretion around the central black hole of the AGN.

Theoretical models for alternate modes to bar-driven fueling also exist. Numerical simulations from \citet{HandQ10} examine several possible mechanisms behind angular momentum transport for a range of galaxy morphologies (bars, spirals, rings, clumpy and irregular shapes, mergers) at different radial scales. For each morphological type, gas transported from larger to smaller ($\sim1$~kpc) radii ``piles up'' due to decreasing efficiency in the processes that induce torque. If this pile-up of gas is sufficiently massive, it becomes self-gravitating and can efficiently transport angular momentum down to scales of $\sim10$~pc. This ``stuff within stuff'' model is similar to the second half of \citet{Shlos89}'s ``bars within bars'' model. The difference is that the ``bars within bars'' model assumes that a large-scale bar is the primary mechanism that transports the gas inward to form the gaseous disc, while \citet{HandQ10} show that many large-scale morphologies are capable of producing a secondary instability and fueling an AGN, suggesting that this process may not be restricted to classic large-scale bars. 

Many studies have focused on observational correlations between the presence of a galactic bar (typically identified at optical wavelengths) and that of an AGN (identified by optical line ratios or widths). Some studies \citep[eg,][]{Ho1997,Mulchaey1997,Hunt1999} find similar bar fractions for both AGN and inactive galaxies and hence report no correlation. The significance of these fractions, however, is hindered by small sample sizes, typically with fewer than 100 barred AGN hosts. More recent studies \citep{Knapen2000,Laine2002,Laurikainen2004} report increases of $20-23$\% in the bar fractions for AGN when compared to non-AGN hosts. Despite larger numbers of AGN, the results are still only significant at the $2.5\sigma$ level. Rather than comparing the likelihood of active and inactive galaxies to host bars, as is most common among previous studies, \citet{Cisternas2013} accounted for a continuum of values by quantifying bar strength and activity level in local X-ray identified AGN. While no correlation was found, these data probe only the low-luminosity AGN regime ($L_X\sim4\times10^{38}$~erg~s$^{-1}$). In the high redshift universe, \citet{Cheung2015} find no compelling evidence that bars are more likely to lie in AGN hosts than non-AGN hosts. 

Several recent studies have focused on optical identifications of bars and AGN, primarily using data from the Sloan Digital Sky Survey (SDSS). We compare these methods and results in Table~\ref{table}. Among these studies, neither \citet{Lee12} nor \citet{Martini03} find any correlation between the presence of strong galactic bars and AGN, but do not rule out the possibility of smaller, nuclear bars influencing AGN activity. In contrast, \citet{Oh12,Hao09,Alonso13} all find evidence of bar effects in AGN --- however, they disagree on both the strength of the effect and whether it affects both black hole fueling and/or central star formation. One possible reason for the discrepancy is the lack of a consistent scheme for classifying AGN. While the BPT diagram based on optical line ratios \citep{BPT} is among the most common methods for identifying AGN, the demarcation between star-forming and AGN host galaxies is not consistent; some use the \citet{Kewley01} criterion that excludes composite galaxies, while others use \citet{Kauffmann03a} and include these along with Seyferts as AGN. The inclusion of LINERs can also complicate the picture; the high line ratios in at least some LINERs are spatially extended and thus likely of a non-AGN origin \citep{Sarzi10,Yan12,Singh2013}. 

Other challenges result from the task of identifying galactic bars, which is often done by visual inspection of optical images by individuals or small groups of experts. This introduces potential complications when there is disagreement between classifiers, especially in the cases of weak or nuclear bars. With only a single (or a few) classifications per image, such disagreements are difficult to resolve. Furthermore, individual visual inspection can limit the effective sample size due to the amount of time required to inspect images one by one. Our work avoids these problems by using crowdsourced citizen science classifications to identify galactic bars, where many individuals (an average of 27~classifiers for bar detection in this study) analyze each galaxy, and the presence of a bar is quantified as a calibrated vote fraction.
   
This paper re-examines the relationship between bars and AGN in disc galaxies by using Galaxy~Zoo morphological classifications, and by using a strict AGN classification scheme which only selects Seyfert galaxies. We use this data to consider three physical scenarios for describing the role bars may (or may not) play in AGN fueling: I) Bars are necessary to fuel AGN, II) Bars are one of several ways to fuel AGN, or III) Bars do not fuel AGN. We discuss each of these possibilities in Section~\ref{sec:Discussion} and suggest the means by which the existence barred AGN, unbarred AGN, barred non-AGN, and unbarred non-AGN may be explained within the context of each model. We then report the scenario which we find to be best supported by both our observations and current theoretical models and simulations.  

In Section~\ref{sec:Sample Selection} we describe our sample selection. Section~\ref{sec:Results} includes our data, with mass and colour distributions of the different activity types, both barred and unbarred, as well as a comparison between accretion strengths of barred and unbarred AGN. Interpretations of these results are discussed in Section~\ref{sec:Discussion}, and the main conclusions are outlined in Section~\ref{sec:conclusions}. We adopt a $\Lambda$CDM cosmology throughout the paper of $\Omega_{m}=0.27$ and $H_0=\rm71~km~s^{-1}~Mpc^{-1}$ \citep{pla13}. 

\begin{table*}
\begin{minipage}{7.5in}
\begin{tabular}{@{}p{2.5cm}p{2.6cm}p{2.2cm}p{2.2cm}p{2.5cm}p{2.5cm}p{2.5cm}@{}}
\hline
\hline
                                      & \citet{Martini03}                     & \citet{Hao09}                          & \citet{Lee12}         & \citet{Oh12}                     & \citet{Alonso13}                 & This work \\
\hline                                                                                                               
Redshift range                        & $z<0.038$                             & $0.01 < z < 0.03$                      & $0.02<z<0.055$        & $0.01<z<0.05$                    & $z<0.1$                          & $0.01<z<0.05$ \\[5mm]
Abs. magnitude range                  & $B_{T}<13.4$                          & $18.5<M_g<-22.0$                       & $M_r<-19.5+5\log(h)$  & $M_r < -19$                      & $M_g<-16.5$                      & $M_{z,petro}<-19.5$ \\[5mm]
Inclination limit                     & $R_{25}<0.35$                         & $i < 60\degr$                          & $b/a>0.6$             &  $b/a > 0.7$                     & $b/a>0.4$                        & \pnotedgeon$>0.6$ \\[5mm]
AGN classification method             & varied                                & $\rm FWHM(H\alpha)>1200~km/s$ and Ka03 & Ke01                  & Ka03                             & Ka03                             & S07, WISE \\[15mm]
AGN type(s)                           & Type 1 and 2 Seyferts, LINERs         & Type 2 Seyfert, LINER, composite       & Type 2 Seyfert, LINER & Type 2 Seyfert, LINER, composite & Type 2 Seyfert, LINER, composite & Type 2 Seyfert \\[10mm]
Bar classification method             & visual inspection                     & ellipse fitting                        & visual inspection     & visual inspection                & visual inspection                & crowdsourced visual inspection \\[5mm]
Number of AGN in sample               & 28                                    & 128                                    & 1742                  & 1397                             & 6772                             & 681\\[5mm]
Fraction of AGN hosts that are barred & 28.6\%                                & 47\%                                   & 49\%                  & 51\%                             & 28.5\%                           & 51.8\% \\[5mm]
\hline
\hline
\label{table}
\end{tabular}
\caption{Summary of recent studies comparing the presence of galactic bars and active galactic nuclei, including new results from this work. \citet{Martini03} is the only study with neither uniform selection criteria for galaxies nor a volume-limited sample. AGN classifications from optical line ratios and the BPT diagram are separated by the following demarcations: Ke01 = \citet{Kewley01}; Ka03 = \citet{Kauffmann03a}; S07 = \citet{Ski07}.}
\end{minipage}
\end{table*}

\section{Data and sample selection}
\label{sec:Sample Selection}

\begin{figure}
\centering
\includegraphics[width=3.5in]{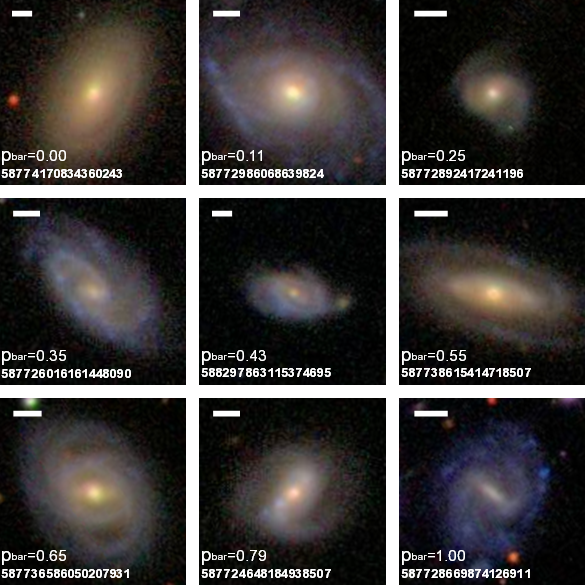}
\caption{Examples of the SDSS images used in \gztwo, sorted by increasing \pbar~(the weighted percentage of users that detected a bar in each image). All galaxies are from our final analysis sample of ``not edge-on'' disc galaxies. The white lines in the upper left of each image represent a physical scale of 5~kpc. We also give \pbar~and the SDSS objectIDs for each galaxy. \textbf{Top row:} Galaxies with \pbar$<0.3$, which in this paper are designated as unbarred. \textbf{Middle and bottom rows:} Galaxies with \pbar$\ge0.3$, which we designate as reliably barred.}
\label{gal}
\end{figure}

Our parent sample of galaxies is taken from the SDSS Data Release 7 \citep{Abazajian09}. From the spectroscopic Main Galaxy Sample \citep{Strauss2002}, we select galaxies within the redshift interval $0.01<z<0.05$ --- the lower limit excludes galaxies whose angular size significantly exceeds the spectroscopic fiber, and the upper limit is chosen so that a reasonable estimate of bar detection can be made by visual inspection. From this, we create a volume-limited sample by applying an additional cut of $M_{z,petro} < -19.5$~AB~mag. 

Within the volume-limited sample, we use morphological cuts to select only disc galaxies at low inclination angles that are candidates for the presence of galactic bars (described below). These cuts result in the final sample of 19,756 disc galaxies used in the remainder of this paper. 

\subsection{Bar classifications and Galaxy~Zoo~2}
To select disc galaxies and measure the presence of a bar, we use data from the online citizen science project \gztwo~(GZ2).\footnote{zoo2.galaxyzoo.org} With the help of over 80,000 volunteers providing over 16~million classifications of over 300,000 galaxies, \gztwo~is the largest extant survey of detailed galaxy morphology. Volunteers are shown colour images of galaxies taken from the SDSS (Figure~\ref{gal}), and are then prompted through a decision tree in which they answer questions about the galaxy's structure. For a detailed discussion on the \gztwo~project and its decision tree, see \citet{Kyle}.

Since bars only appear in disc galaxies, the sample must be limited to disc galaxies in which a bar can be seen via visual inspection. We begin by selecting galaxies for which at least 10~people answered the question, ``Is there a sign of a bar feature through the centre of the galaxy?'', thus rejecting vote fractions with low statistical significance. Because questions in GZ2 are implemented as part of a decision tree \citep{Kyle}, users must have identified a galaxy as a disc and as not edge-on before answering the bar question. In this way, the cut of $N_{bar}\ge10$ increases the likelihood that the galaxy in question is a candidate for having a bar. This cut is not complete, however, for galaxies which have a high number of total classifications. In these cases, the number of users to answer the bar fraction may still be small compared to the number of users identifying the galaxy as either not disc-like, or as an edge-on galaxy. Therefore cuts are also applied to the vote fractions relating to questions preceding the bar question. The first question of the GZ2 tree reads, ``Is the galaxy simply smooth and rounded, with no sign of a disc?'' \citet{Kyle} determined the threshold fraction of ``features or disc'' answers required to classify the galaxy as a disc, when combined with the cut $N_{bar}\ge10$, to be \pfeatures$\ge0.227$. We emphasize that the cuts provided in \citet{Kyle} are intended to be \emph{minimum} values for determining well sampled galaxies. We thus chose to adopt a slightly higher value of \pfeatures$\ge0.35$ to create the cleanest possible sample, based on a visual inspection of a subsample of galaxies with these cuts. To assess whether the results would be affected by this choice, we also created a sample with the original \citet{Kyle} cuts. This choice increased the number of AGN in the sample by 24, and did not affect the final results. Therefore we present the sample using our more conservative cuts in this paper. 

Following an answer of ``features or disc'' for the first question, the volunteer is then asked ``Could this be a disc viewed edge-on?'' Bars become increasingly difficult to detect in galaxies at high inclination angles, and are nearly impossible to detect in edge-on galaxies without careful isophotal mapping. The threshold vote fraction determined by \citet{Kyle} of a ``No'' answer to this question is \pnotedgeon$\ge0.519$. We again adopt a slightly more conservative value of \pnotedgeon$\ge0.6$ based on visual inspection of a subsample. The combination of feature/disc galaxies that are not edge-on for these two thresholds results in the final sample size of 19,756~galaxies used in this paper (Table~\ref{tab:cat}). 

\begin{figure*}
\includegraphics[width=7in]{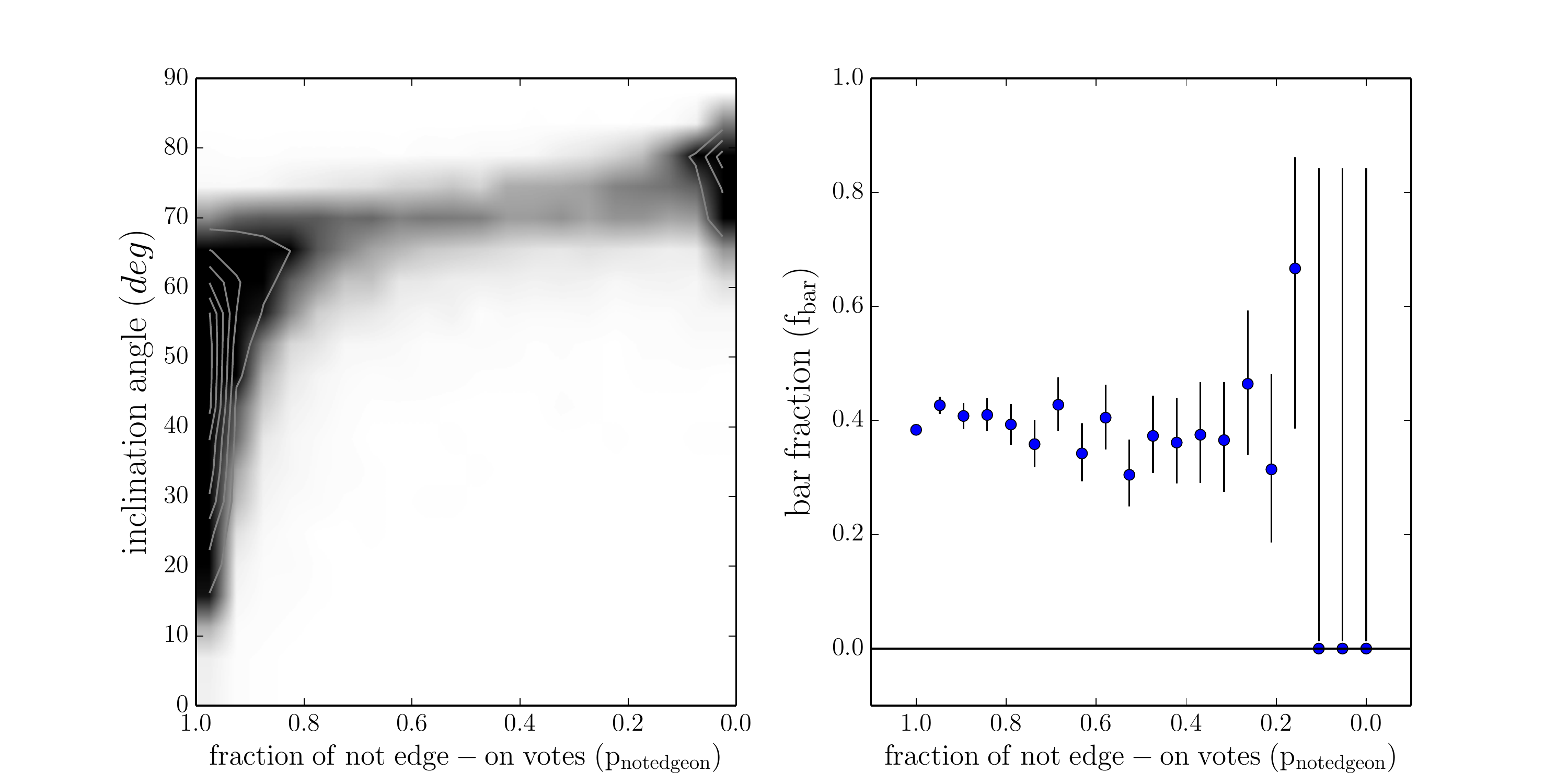}
\caption{Left: Fraction of ``not-edge-on'' votes vs. inclination angle ($i = \cos^{-1}[a/b]$) for the disc galaxies in our GZ2 sample. An angle of $0\degr$ means the galaxy is completely face-on, while $90\degr$ is completely edge-on. GZ2 users consider a galaxy as ``not edge-on'' if the inclination angle is less than $i\sim70\degr$. Right: Fraction of barred galaxies vs. fraction of ``not edge-on'' galaxies. The bar fraction is independent of the edge-on degree of the galaxies (above $p_{notedgeon}\sim0.3$); the ability of users to detect bars does not decrease with inclination until $p_{notedgeon}\sim0.3$, or $i\sim70\degr$. Error bars are 95\% Bayesian binomial confidence intervals \citep{Cameron11}. This demonstrates that GZ2 data can reliably identify bars even in moderately-inclined disc galaxies.}
\label{angles}
\end{figure*}

As a check that our selection of ``not edge-on'' disc galaxies can be reliably used to identify a bar, we examine the inclination angle of the sample, which is approximated by the ratio of the best fit of the semi-major and -minor axes $i = \cos^{-1}(a/b)$ as measured in $r$-band by the SDSS pipeline. Figure~\ref{angles} shows the strong correlation between $i$ and \pnotedgeon, with a sharp cutoff near $i=70\degr$. Our cutoff of \pnotedgeon$\ge0.6$ effectively limits the sample to inclination angles of $i<67\degr$. In Figure~\ref{angles} we also show the dependence of the GZ2 bar fraction on \pnotedgeon. The bar fraction remains roughly constant ($\pm0.1$) between $0.3<$\pnotedgeon$<1.0$ and drops to zero at \pnotedgeon$<0.1$. Since the \emph{true} bar fraction is expected to be independent of $i$ (a purely geometrical effect assumed to have a random distribution), any change in the bar fraction would reflect the ability of visual inspection to detect a bar in a highly inclined disc. The constant bar fraction out to our limit of \pnotedgeon$\ge0.6$ (and well beyond) is a necessary requirement for an unbiased selection of barred galaxies; as a result, we are confident that the crowdsourced bar classifications in this sample are reliable.

Finally, if the volunteer answers ``No'' to the edge-on question, they are asked ``Is there a sign of a bar feature through the centre of the galaxy?'' Possible answers to this question are either ``Bar'' or ``No bar''. \citet{Kyle} compared expert classifications of barred galaxies from both \citet{NA10} and \citet{EFIGI} to \gztwo~data, and show that a threshold of \pbar$\ge0.3$ is the most reliable separator of the barred from unbarred population (see their Figure~10). We adopt the same threshold of \pbar$\ge0.3$ for determining whether a galaxy has a bar (see Figure~\ref{gal} for images of galaxies with different values of \pbar).

We compare our morphology cuts to those used by \citet{Masters11}, who used an early release of GZ2 data to identify barred galaxies. Their study also required $N_{bar}\ge10$ and claim that this cut alone is sufficient to restrict the sample to disc galaxies without applying an additional cut on \pfeatures. This assumption was reasonable at the time since the \gztwo~project was still collecting data, and the number of classifications per galaxy was lower than in the final catalog. The median number of classifications per galaxy is roughly 30\% higher, and so our data is more susceptible to contamination by non-disc galaxies with high classification counts. This makes an additional cut on \pfeatures necessary. To remove edge-on discs, \citet{Masters11} set an inclination limit of $\log(a/b)<0.3$, or $i\sim60\degr$; this is comparable to our \pnotedgeon cut, which corresponds to roughly $i\sim67\degr$. To select barred galaxies, a majority vote fraction of \pbar$>0.5$ was required, higher than our value of \pbar$\ge0.3$. We are nevertheless confident in our threshold which was determined by the more recent and detailed analysis of the GZ2 data by \citet{Kyle} as described above. Additionally, the data released at the time of \citet{Masters11} had not yet been reduced via weighting and debiasing; these differences in vote fractions also contribute to the different cuts used in our study. 

\subsection{Activity type classification}
\label{sec:Activity}
\begin{figure*}
\includegraphics[width=7in]{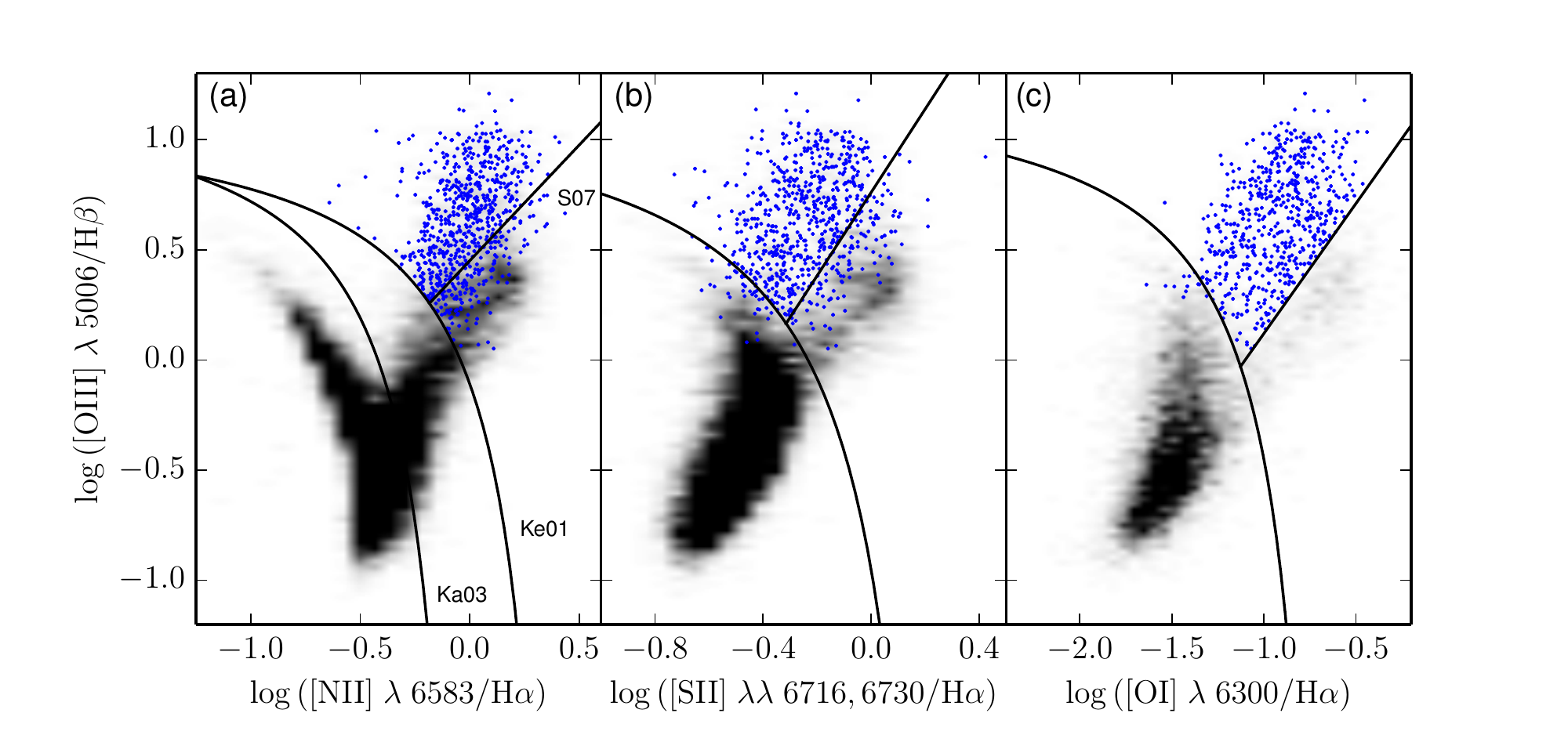}
\caption{Optical line diagnostics for activity types of 19,756 disc galaxies. Any galaxy with $\rm S/N < 3$ for \ion{O}{iii}, H$\beta$, \ion{N}{ii}, or H$\alpha$ is unclassifiable using this method and labeled as ``undetermined''. The 3,619 undetermined galaxies do not appear on the diagram above. The remaining 16,137 galaxies were categorized according to the above diagrams in the following order, based on the method of \citet{Ski07}. First, diagram (a) was used to identify star-forming and composite galaxies. Any galaxy below the Ka03 line was classified as star-forming, while those that fell between the Ka03 and Ke01 lines were classified as composite. Next, to distinguish AGN from LINERs, we use diagrams (b) and (c). If a galaxy had $S/N > 3$ for \ion{O}{i}, diagram (c) was used. If a galaxy did not have $S/N > 3$ for \ion{O}{i}, but did for \ion{S}{ii}, diagram (b) was used. Last, if a galaxy did not have $S/N > 3$ for \ion{O}{i} or \ion{S}{ii}, but did for \ion{N}{ii}, diagram (a) was used. In each panel, only galaxies with $S/N > 3$ for all four lines required by that diagram are shown. Galaxies designated AGN by any of the three optical line diagnostics are plotted as blue points, while the black shading represents the full sample of emission-line galaxies.}
\label{BPT}
\end{figure*}

We use flux measurements from the 2012 release of the OSSY catalogue \citep{OSSY} to classify disc galaxies as either star-forming, composite, AGN, LINER, or quiescent (also known as ``undetermined''). This method employs ratios of \ion{O}{iii}/$\rm H\beta$ fluxes as a function of \ion{N}{ii}, \ion{S}{ii}, or \ion{O}{i} over $\rm H\alpha$ according to the BPT diagnostics. Our method for selecting AGN is the same as used by \citet{Ski07,Ski10}. First, we use the \ion{N}{ii}/H$\alpha$ ratio (Figure~\ref{BPT}a). Any galaxy that does not have $\rm S/N>3$ for any of the four lines is unclassifiable via this method (possibly due to being gas-poor) and labeled ``undetermined.'' Next, any galaxy which falls below the \citet{Kewley01} extreme starburst line is classified as star-forming, and those that fall between this and the \citet{Kauffmann03a} empirical starburst line are classified as composite. We note that some of these composite galaxies may be potential AGN, but we cannot cleanly separate the AGN contribution from star formation and thus exclude them from our sample \citep{Ski10}. 

Next, we identify the remaining galaxies (above the extreme starburst line) as either Seyfert AGNs or LINERs. \citet{Kewley06} showed that both \ion{O}{i}/H$\alpha$ and  \ion{S}{ii}/H$\alpha$ diagrams are better-suited to distinguish AGN from LINERs; we thus use diagram (c) in Figure~\ref{BPT} if these galaxies also have $\rm S/N>3$ in \ion{O}{i}. For galaxies which do not have $\rm S/N>3$ in \ion{O}{i}, but do in \ion{S}{ii}, we use diagram (b). In both cases, we use the AGN-LINER division line of \citet{Kewley06}. For the remaining galaxies, we use diagram (a) and implement the AGN-LINER division line of \citet{Ski07}.

Finally, to detect any AGN that may have been optically mis-classified due to obscuration, we identify AGN based on their infrared continuum shape using data from the Wide-field Infrared Survey Explorer \citep[WISE]{WISE}. We identify as an AGN any galaxy with $(W1-W2) \ge 0.8$ \citep{Stern12}. Based on infrared data, we re-classified fourteen galaxies (originally classified optically as three star-forming, ten composites, and one LINER) as AGN. 

We show the results of the activity type and morphological classifications in Table~\ref{activitytable}. The numbers and fractions of each activity type with respect to the full sample are shown, as well as the numbers and fractions of barred galaxies within each activity type. These results are discussed in Section~\ref{sec:Results}. 

\begin{table}
\begin{center}
\begin{tabular}{lcccccc}
\hline
\hline

 &  \multicolumn{2}{c}{All discs} & & \multicolumn{2}{c}{Barred discs} \\
\cline{2-3} \cline{5-6}
Activity type & Number & $f_{\rm total} (\%)$  & & Number & $f_{\rm bar} (\%)$   \\
\hline

star-forming    & 11282 & 57.1  $ \plusminus{+0.7}{-0.7} $&     & 4183  &   37.1  $ \plusminus{+0.9}{-0.9} $  \\
composite       & 2853  & 14.4  $ \plusminus{+0.6}{-0.4} $&     & 1301  &   45.6  $ \plusminus{+1.8}{-1.8} $  \\
AGN             & 681   &  3.4  $ \plusminus{+0.3}{-0.2} $&     & 353   &   51.8  $ \plusminus{+3.8}{-3.7} $  \\
LINER           & 1321  &  6.7  $ \plusminus{+0.4}{-0.4} $&     & 695   &   52.6  $ \plusminus{+2.7}{-2.7} $  \\
undetermined    & 3619  & 18.3  $ \plusminus{+0.6}{-0.5} $&     & 1654  &   45.7  $ \plusminus{+1.6}{-1.6} $  \\
\hline
total           & 19756 & 100 &       & 8186  &   41.4  $ \plusminus{+0.7}{-0.7} $  \\
\hline

\end{tabular}
\caption{Results of activity classification for our sample of 19,756 not edge-on disc galaxies. $f_{\rm total}$ is the percentage of the total sample represented by each activity (number of galaxies of that type / total number of galaxies). $f_{\rm bar}$ is the percentage of each subsample that are barred (number of galaxies of that type that are barred / total number of galaxies in that type). Errors are 95\% Bayesian binomial confidence intervals \citep{Cameron11}.}
\label{activitytable}
\end{center}
\end{table}

\section{Results}
\label{sec:Results}

To determine whether a correlation exists between galaxies that host an AGN and those that contain large-scale stellar bars, we examine the fractions of barred and unbarred AGN with respect to mass, colour, and AGN strength. We use stellar masses from the {\tt AVERAGE} values in the MPA-JHU DR7 catalogue \citep{Kauffmann03b}. Colours are $^{0.0}(u-r)$ values from SDSS DR7, which have been both de-reddened for Galactic extinction and $k$-corrected to redshift $z=0.0$ \citep{csa03}. Stellar velocity dispersions are taken from \citet{OSSY}. An excerpt of these data may be found in Table~\ref{tab:cat}.

\subsection{Barred AGN fraction at a fixed mass and colour}\label{ssec:barredfraction}
 
\begin{figure*}
\includegraphics[height=3.8in]{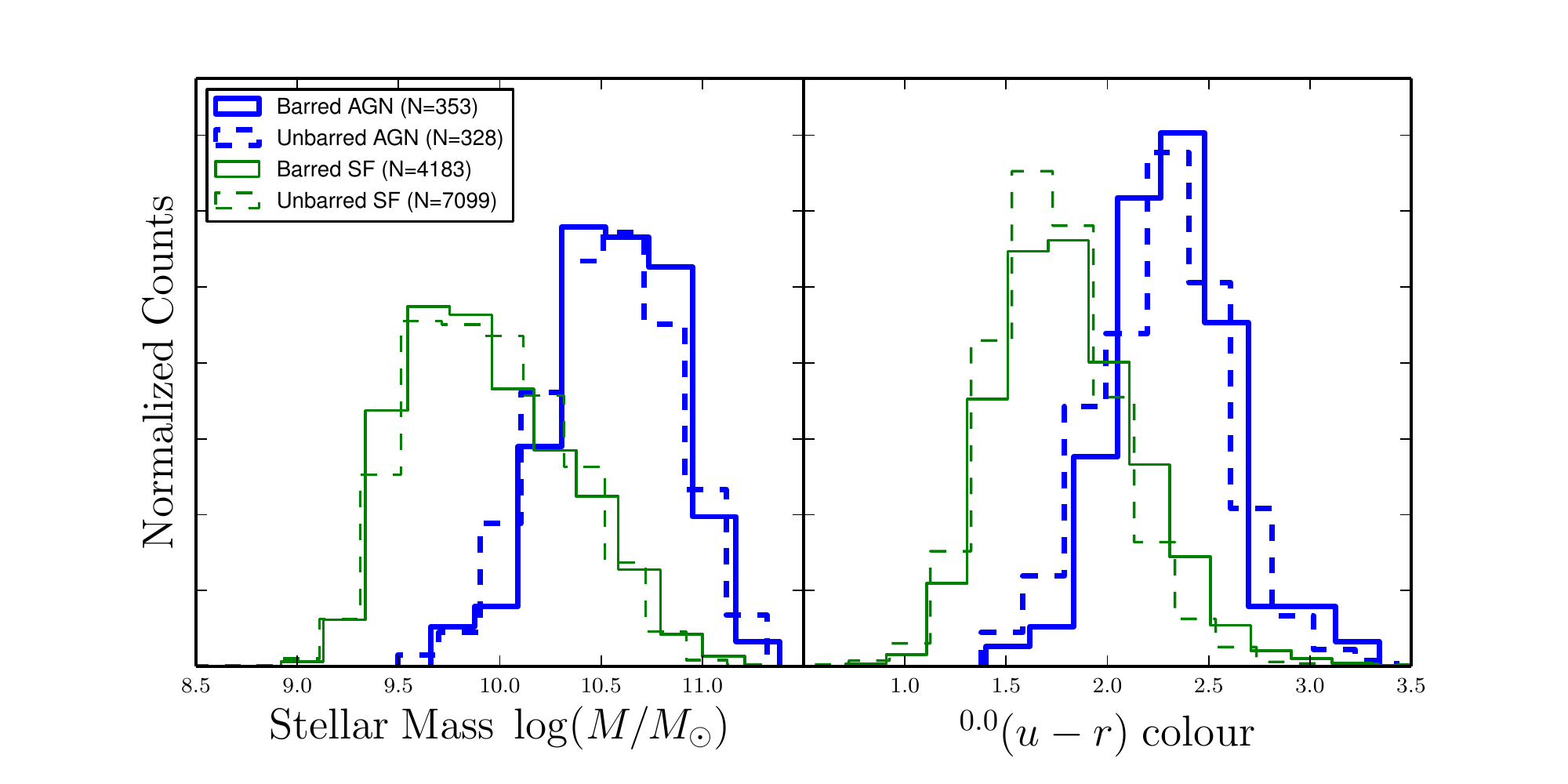}
\caption{Mass and colour distributions for disc galaxies in the GZ2 sample, separated by both activity type (either AGN or star-forming as in Table~\ref{activitytable}) and the presence of a galactic bar. AGN (green) are on average both significantly redder and more massive than star-forming galaxies (blue). When splitting the disc galaxies into barred (solid lines) and unbarred (dashed lines), however, there is no significant difference between the two populations. Counts are normalized so that the sum of bins is equal to 1 for each sample.}
\label{AGNSFHistogram}
\end{figure*}

Figure~\ref{AGNSFHistogram} shows the distributions of mass and colour for AGN and star-forming activity types, split into barred and unbarred subsamples. The median AGN is more massive (by $0.6$~dex) and redder (by $0.5$~mag) than the median star-forming galaxy. This agrees with previous optical studies of AGN and star-forming galaxies in the local Universe \citep{Ski07,Lee12,Oh12,Alonso13}. \citet{Aird12} demonstrate that this difference is primarily caused by selection effects relating to the underlying Eddington ratio distribution. The probability of a galaxy hosting an AGN is assumed to be independent of stellar mass, and thus AGN are prevalent at all masses in the range $9.5<\log(M/M_{\odot})<12$, despite only being observable at higher masses. As a result, we expect higher absolute numbers of barred AGN in a flux-limited sample since barred disc galaxies are also on average redder and more massive than unbarred disc galaxies \citep{Masters11,Masters12}. We interpret this as the primary cause for the higher fraction of barred AGN (51.8\%) versus barred star-forming (37.1\%) galaxies in Table~\ref{activitytable}.

\begin{figure*}
\includegraphics{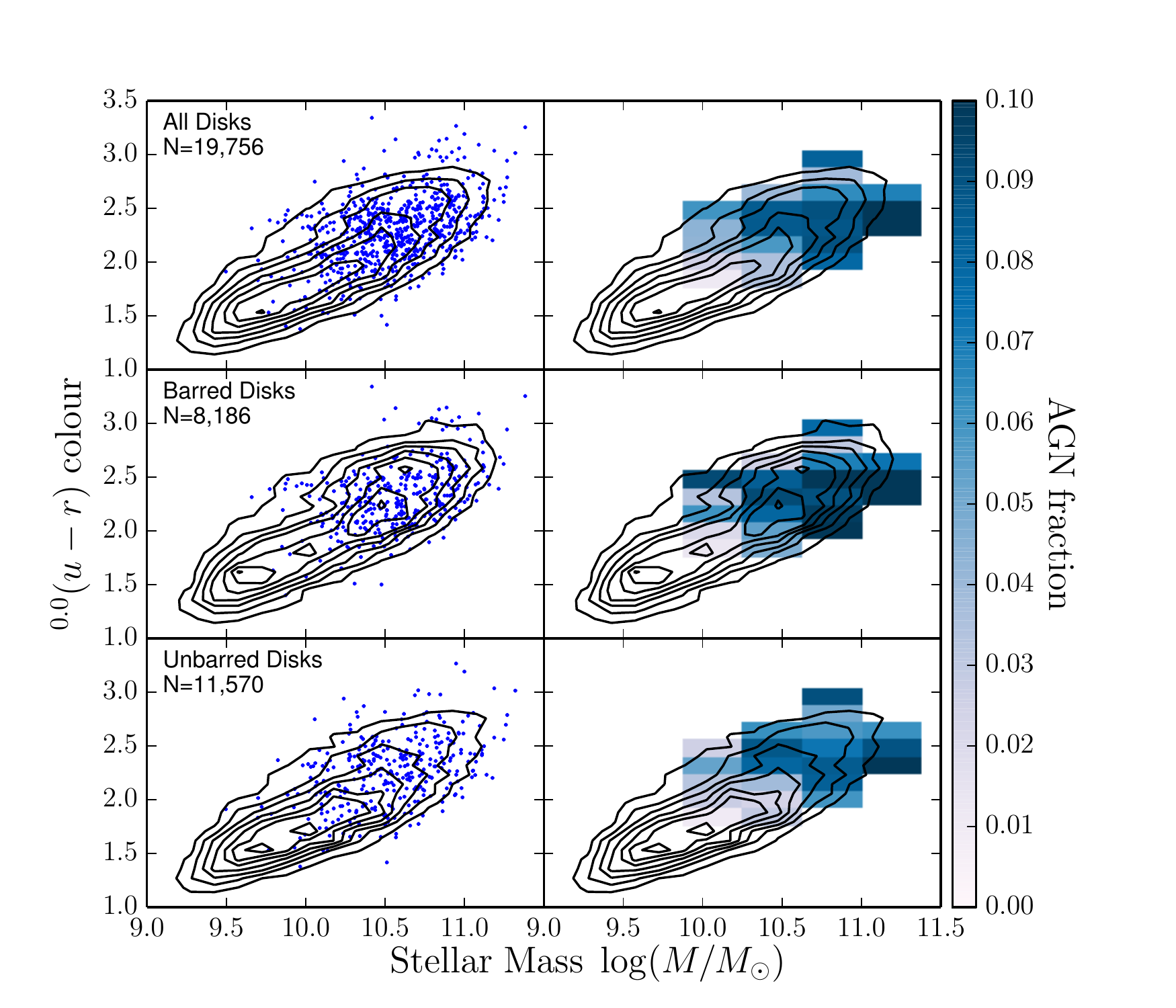}
\caption{Optical colour vs. stellar mass for disc galaxies in GZ2. Black contours represent all disc galaxies (top), all barred galaxies (middle), or all unbarred galaxies (bottom). All AGN (top), barred AGN (middle), and unbarred AGN (bottom) are plotted in the left panels as blue dots; the right panels show the AGN fraction in each colour/mass bin. Bins with $N_{AGN}<10$ are masked.}
\label{BlueMasscolor}
\end{figure*}

To control for this selection effect, we examine the fraction of AGN at fixed masses and colours (Figure~\ref{BlueMasscolor}). The total disc galaxy sample spans a mass range from $9.0<\log(M/M_\odot)<11.5$, while the colour range extends from $1.0<(u-r)<3.5$. AGN hosts are found throughout the disc galaxy sample, but most appear in galaxies with $\log(M)>10^{10}~M_\odot$. When examining the fraction of galaxies with an AGN as a function of mass and colour, redder and more massive galaxies have AGN fractions as high as 10\%. Bins with fewer than 10 total AGN (barred AGN + unbarred AGN) are masked to minimize variance from small sample sizes. The same trend is also seen when splitting the disc galaxy sample into barred and unbarred subsamples. 

\begin{figure*}
\includegraphics{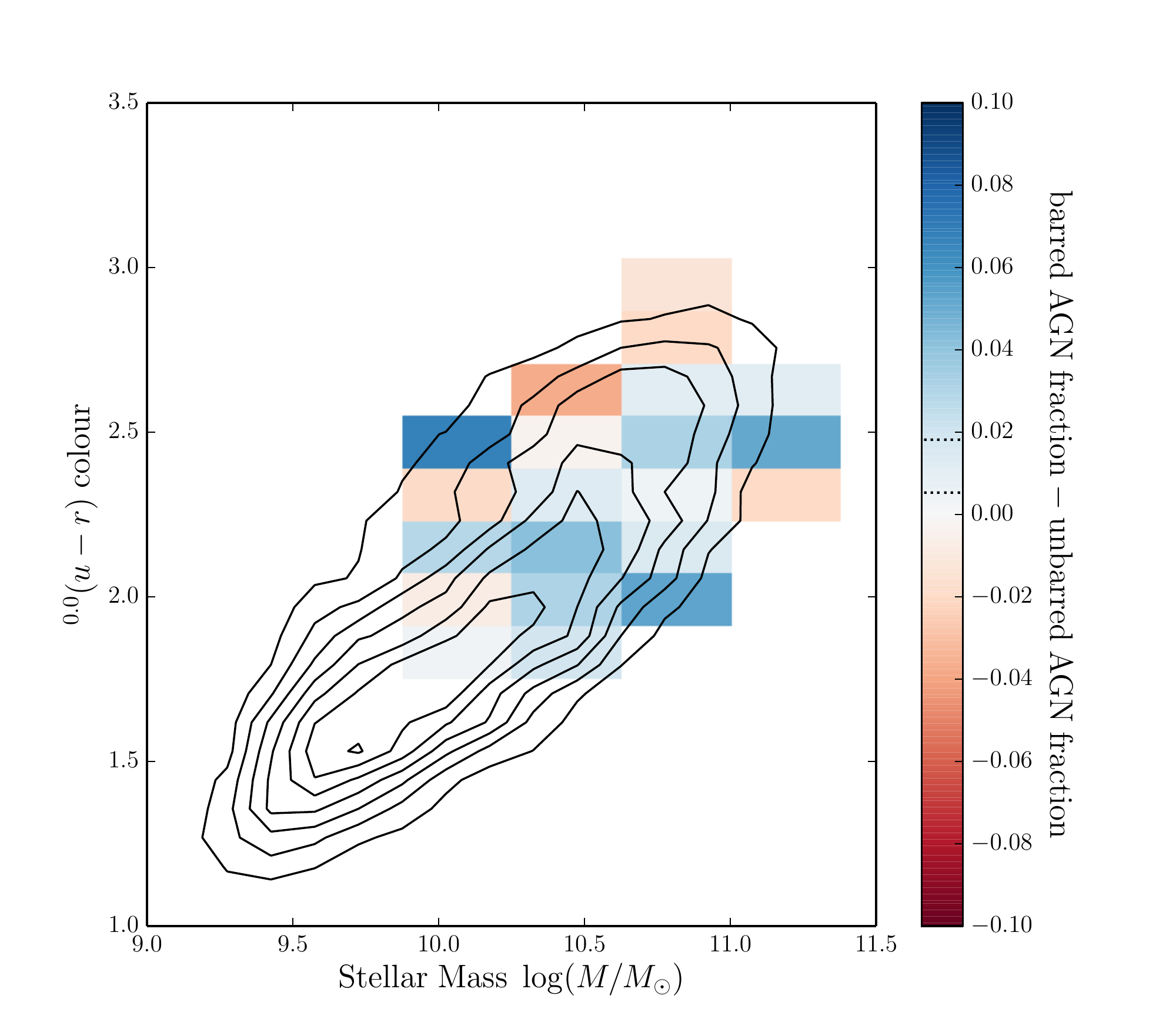}
\caption{Optical colour vs. stellar mass for barred and unbarred disc galaxies in GZ2. Coloured bins show the difference between the AGN fractions for barred and unbarred galaxies. Blue bins have higher fractions of barred galaxies, red bins have more unbarred galaxies, and pale/white indicates no difference. The region on the colourbar enclosed by the dotted lines represents the mean of the data determined by the Anderson-Darling test. The colour gradient is on the same scale as Figure~\ref{BlueMasscolor}. Bins with $N_{AGN}<10$ are masked. A colour version of this plot may be found in the electronic edition of the journal.}
\label{RedBlueHist}
\end{figure*}

To analyze the difference between the barred and unbarred AGN populations, we plot the difference in barred and unbarred AGN fractions in Figure~\ref{RedBlueHist}. This quantity is defined as:

\begin{equation}
d_{\rm{B-NB}}=\rm{barred~AGN~fraction - unbarred~AGN~fraction} 
\end{equation}

\noindent and is calculated in each of the mass/colour bins in Figure~\ref{BlueMasscolor}. For each bin, a positive value represents a greater fraction of barred AGN and is coloured blue; a negative value represents a greater fraction of unbarred AGN and is coloured red.

Since our AGN sample is divided into relatively small subsamples, we examine how the size and placement of the mass/colour bins affect the results of Figure~\ref{RedBlueHist}. To control for this effect, we examine the average value of \db~and the fraction of bins with \db~$>0$, defined as:

\begin{equation}
f_{\rm{B>NB}}=\rm{\frac{number~of~bins~with~higher~barred~AGN~fraction}{total~number~of~bins}}.
\end{equation}

\begin{figure}
\centering
\includegraphics[width=3.6in]{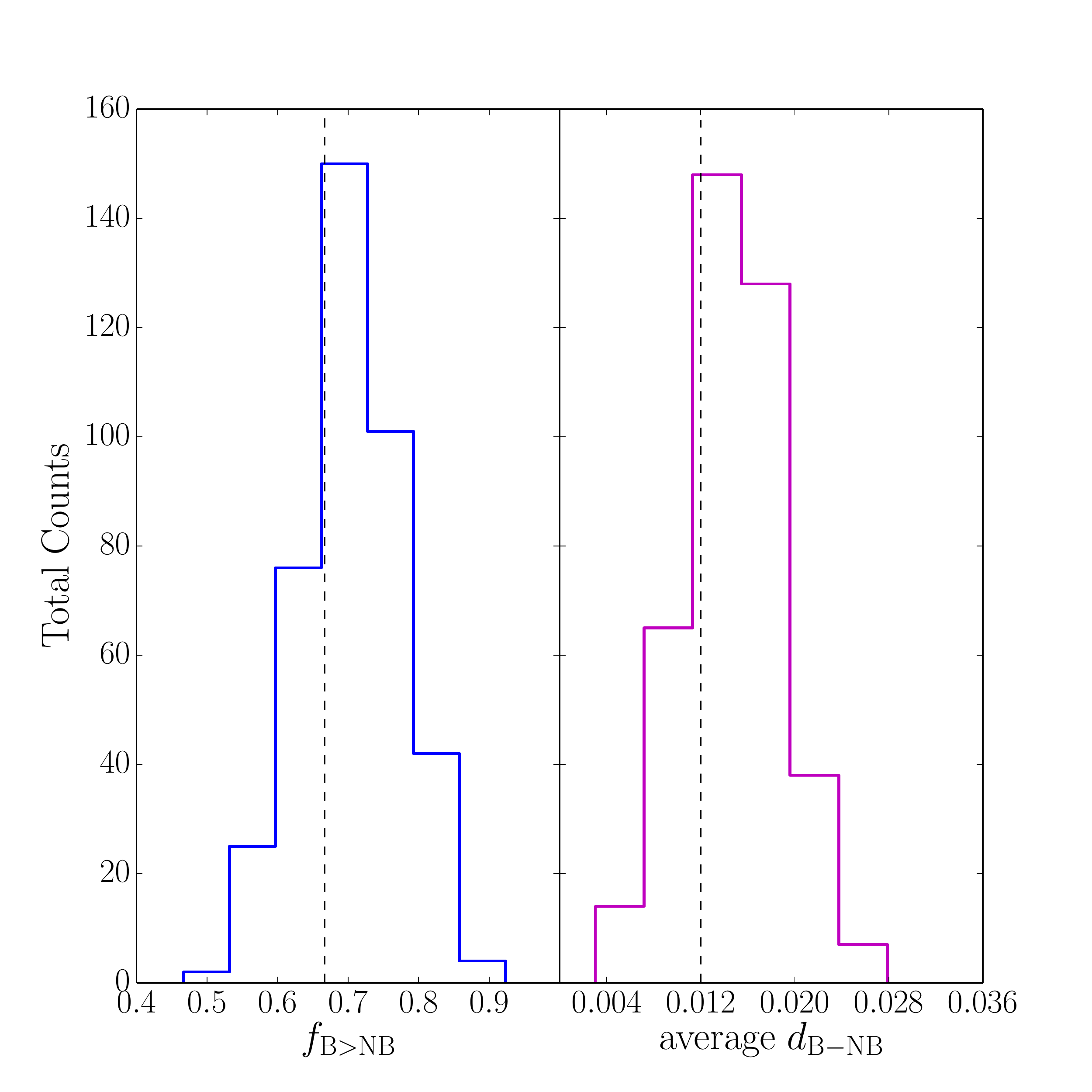}
\caption{Distributions of the difference in the fraction of bins with excesses of barred AGN (\fb) and the average difference between barred and unbarred AGN fractions (\db). Both values are computed for 400~variations in the mass and colour bin widths. \textit{Left}: The average fraction of bins with a higher barred AGN fraction is \fb~$=0.705\pm0.073$. \textit{Right}: The average difference in barred and unbarred AGN fractions is \db~$=0.015\pm0.004$. Dashed black lines indicate the values of \fb~and average \db~used in Figure~\ref{RedBlueHist} and subsequent analysis.}
\label{BinStats}
\end{figure}

\noindent We compute \fb~for 400~combinations of mass and colour bin widths between $0.2 \le \Delta \log(M/M_{\odot}) \le 0.6$ and $0.12 \le \Delta (u-r) \le 0.35$. The distribution of results from all combinations is shown in Figure~\ref{BinStats}. Our final bin choice (as seen in Figure~\ref{RedBlueHist}) has a mass width of $\Delta \log(M/M_{\odot})=0.375$ (16~bins) and colour width of $\Delta (u-r)=0.16$ (22~bins). This choice lies near the peak of the distributions for both \fb~and \db, while maximizing the total number of bins to decrease the uncertainty on statistical tests. 

For the first time among recently published studies, we quantify the level of correlation between the presence of a bar and AGN through statistical analysis. We test the null hypothesis that in the absence of a causal link, the difference between barred and unbarred AGN fractions when binned by mass and colour should be centered around zero. The null hypothesis also requires that the likelihood distribution decreases symmetrically from zero in both directions; as a result, we assume a normal distribution with mean $\mu=0$ and standard deviation $\sigma$. Other models of the null hypothesis could of course also be tested, but we adopt this as the simplest reasonable scenario that fits the constraints of the problem. 

\begin{figure*}
\centering
\includegraphics[width=7.0in]{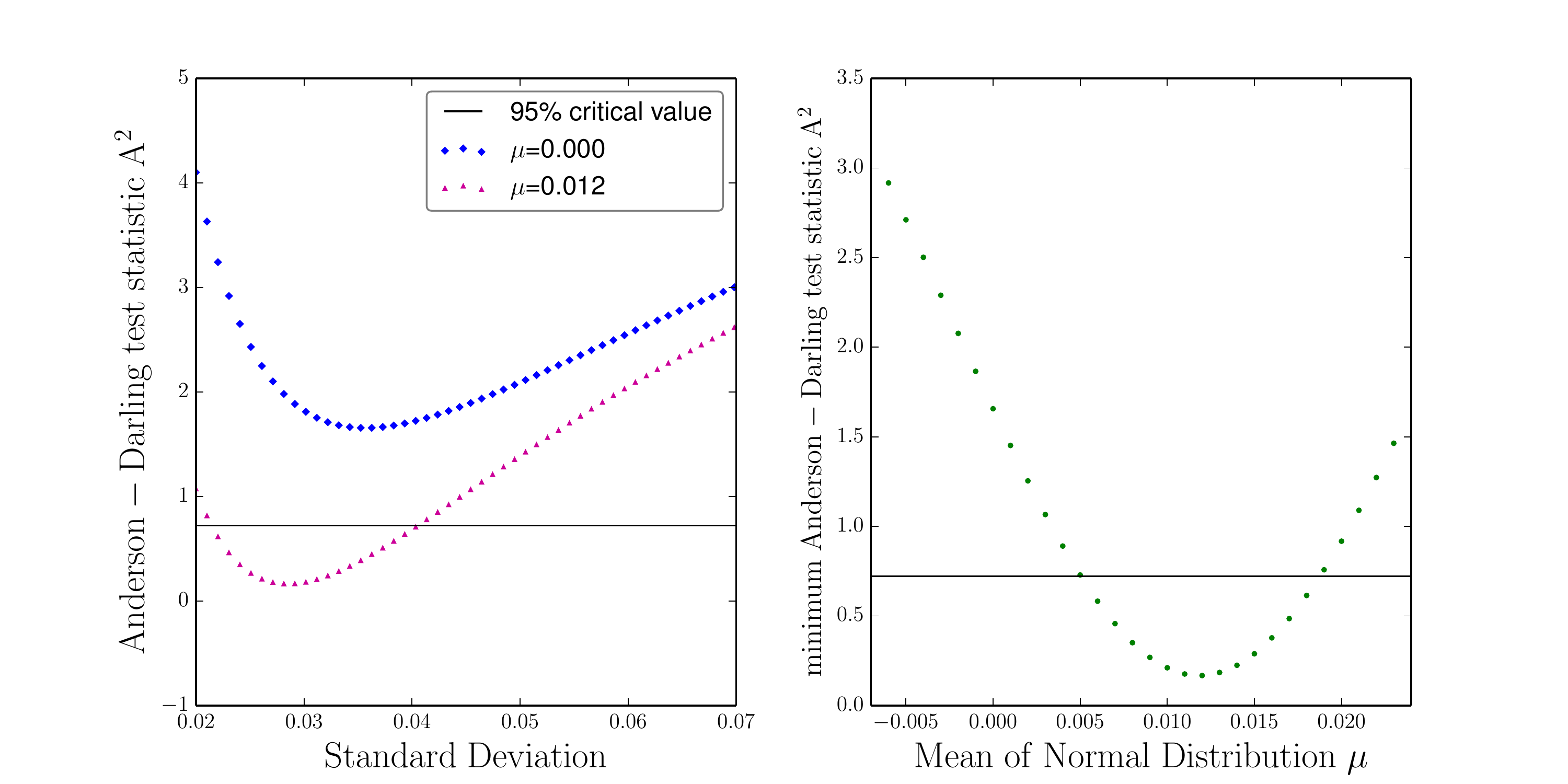}
\caption{Fits of the binned fraction of barred vs. unbarred AGN fractions to a normal distribution. \textit{Left:} value of the Anderson-Darling test ($A^2$) as a function of the standard deviation of the normal distribution being fit ($\sigma_d$). The horizontal black line shows the critical value of $A^2$ corresponding to 95\%; a model must fall below this line to be considered an acceptable fit at this level of confidence. Two models are shown: the null hypothesis (blue diamonds) and the best fit to the data in Figure~\ref{RedBlueHist} (purple triangles). \textit{Right}: Plot of the minimum $A^2$ for the full range of means (\db) tested for the data. This shows that acceptable fits can be found for $0.005<$\db$<0.019$, but that the null hypothesis is ruled out at 95\% confidence.}
\label{ADtest}
\end{figure*}

To assess the level of statistical significance, we fit the data in Figure~\ref{RedBlueHist} with a range of models with varying mean (\db) and standard deviation ($\sigma_d$) and then apply an Anderson-Darling test. We selected this test because it has been empirically shown to be more powerful and reliable at testing normality than traditional $\chi^2$ or Kolmogorov-Smirnov tests, especially with small ($n<30$) sample sizes \citep{Hou2009}. The confidence threshold required for the model to pass at fitting the data is 95\%. In Figure~\ref{ADtest}, we show the distribution of the Anderson-Darling statistic $A^2$ as a function of $\sigma_d$ for two  of the tested models: the null hypothesis (\db~$=0$) and the best fit to the data (\db~$=0.012$). The null hypothesis fails the Anderson-Darling test for all values of $\sigma_d$, indicating that the $66.7\%\plusminus{+16.1\%}{-21.6\%}$ fraction of bins that have a higher barred than unbarred AGN fraction is statistically significant. The best fit to the data, by contrast, has a mean of \db~$=0.012\plusminus{+0.007}{-0.007}$ and $\sigma_d=0.028$. The positive value of \db~indicates an increase in the AGN fraction for \underline{barred galaxies}, consistent with the hypothesis that at least some fraction of AGN activity is triggered or sustained by bar-driven fueling.

\subsection{Comparing barred and unbarred AGN accretion strengths}
\label{sec:Eddington Ratio}
If the presence of a bar does contribute to AGN fueling, one possible result would be an increase in the accretion rate for barred AGN hosts vs. those that are unbarred. To assess this, we compare relative accretion strengths using the quantity $R=L_{\rm{[O~III]}}$/\mbh, with $L_{\rm{[O~III]}}$ as a proxy for the AGN bolometric luminosity. \ion{O}{iii} luminosities were calculated using fluxes from \citet{OSSY}, and black hole masses estimated using the \mbh-$\sigma$ relation:

\begin{equation}
\log \left(\frac{M_{\rm{BH}}}{M_{\sun}} \right) = \alpha + \beta \log \left( \frac{\sigma}{200~{\rm km~s}^{-1}} \right).
\end{equation}

\noindent Here $\alpha$ and $\beta$ are empirical values determined from the observed relationship between black hole mass and velocity dispersion $\sigma$. We adopt the parameters measured by \citet{Gultekin09} of $(\alpha,\beta) = (8.12\pm0.08,4.24\pm0.41)$.  

It has been demonstrated for smaller samples of galaxies that the parameters $\alpha$ and $\beta$ vary as a function of morphological type \citep{Graham11,Gultekin09,Brown13}, including differences between barred and unbarred galaxies. We choose \underline{not} to use $(\alpha,\beta)$ parameters where $(\alpha,\beta)$ are derived from separate subsamples for two reasons. First, since the \mbh-$\sigma$ relation is calibrated from small samples of nearby galaxies, the statistical error on the parameters increases as galaxies are divided into smaller sub-groups. The calibration of \citet{Gultekin09}, for instance, is based on measurements of only eight barred galaxies. The error in $\beta$ for the barred \mbh-$\sigma$ relation is $\sigma_\beta=\pm0.751$, almost twice the error obtained by fitting to the full sample of disc galaxies. Second, while different studies report consistent values for $\alpha$ and $\beta$ when all disc galaxies are considered, the values can vary significantly when splitting by morphological type. \citet{Lee12} and \citet{Alonso13} use separate values for $(\alpha,\beta)$ and report conflicting levels of agreement, depending on which parameters are used. This raises the possibility that differences in AGN strength are simply due to differences in calibration parameters, and not in the true distribution of accretion efficiencies.

\begin{figure*}
	\includegraphics{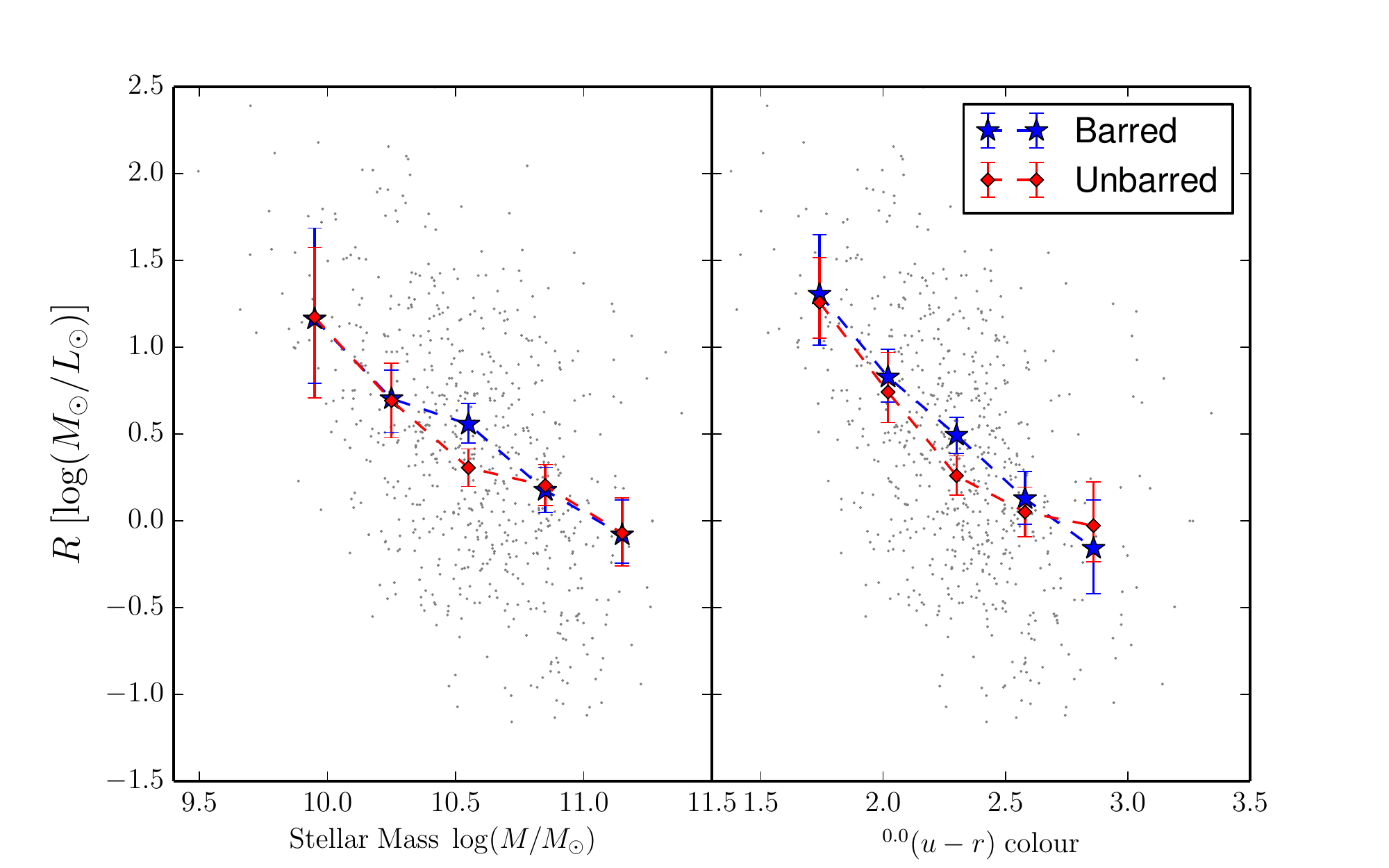}
\caption{\textbf{Left:} Relative accretion strength $R$ vs stellar mass for barred (blue) and unbarred (red) AGN in our sample. $R$ is plotted as the mean of values within five equal-width bins in the range $9.8<\log(M/M_{\odot})<11.3$, which includes 98\% of the AGN sample. Points are drawn at the midpoint of each bin. \textbf{Right:} $R$ vs colour for barred and unbarred AGN. $R$ is plotted as the mean of values within five equal width bins in the colour range $1.6<(u-r)<3.0$, which includes 96\% of the AGN sample. Error bars for each plot are 95\% confidence intervals, calculated by bootstrapping with 1000~times resampling. There is no significant difference in accretion strengths for barred and unbarred AGN as a function of either mass or colour.}
\label{Rate}
\end{figure*}

Figure~\ref{Rate} shows the relative accretion strengths $R$ for our sample as a function of mass and colour for both barred and unbarred AGN; these values are inversely correlated with both mass and $(u-r)$ colour. This trend is likely driven by the same selection effects described in \S\ref{ssec:barredfraction} \citep{Aird12}. At a fixed $L_{\rm{[O~III]}}$/\mbh~ratio, AGN with lower mass black holes are less likely to be detected due to the signal to noise requirements on their spectral lines. This biases the distribution of $R$ toward higher mass black holes. Since stellar mass is strongly correlated with black hole mass \citep{har04,Gultekin09,mer10}, and stellar mass correlates with optical colour \citep{Kauffmann03b}, this explains the trend seen in both parameters for an uncorrected sample. 

Since these observationally-driven selection effects are likely to affect barred and unbarred galaxies equally, we compare the values of $R$ of both groups without any corrections. A two-sided KS-test yields a $p$-value of $p=0.127$ for the two distributions. This is consistent with both the barred and unbarred galaxies being drawn from the same distribution. We thus conclude that there is no strong evidence for a difference in accretion strength between barred and unbarred AGN.  

This result contradicts \citet{Alonso13}, who found an excess of barred AGN with high values of $R$. We conjecture that this may be the result of their sample selection, which excluded galaxies with $M_\star<10^{10}M_\odot$ in favor of a higher redshift limit of $z=0.1$. However, low mass galaxies have higher $L_{\rm{[O~III]}}$/\mbh~ratios and are more likely to be unbarred than their higher mass counterparts \citep{Lee12}. If this effect is real, it appears to be limited to high-mass galaxies (which themselves are subject to selection effects due to the methods used to measure $R$). Additionally, \citet{Alonso13} include composites and LINERs in their sample of AGN. If the activity from these galaxies is not primarily from black hole accretion, $R$ is not a true proxy for accretion strength, and comparisons between barred and unbarred galaxies do not accurately probe differences between the two populations. To test this, we compare $R$ distributions for barred and unbarred composite + AGN + LINER galaxies with $M_\star>10^{10}M_\odot$. For these galaxies, the difference in the average values of $R$ for the barred and unbarred samples is 0.09~$(L_\odot/M_\odot)^{-1}$ (compared to a difference of 0.06~$(L_\odot/M_\odot)^{-1}$ when considering only AGN with no cut on stellar mass), and a KS-test for the distributions yields a p-value $<0.01$, which agrees with the results of \citet{Alonso13}. We note that our results are consistent with \citet{Lee12}, who have a similar mass range to our sample of disc galaxies, and do not include composites in their sample.

\section{Discussion}\label{sec:Discussion}

We have compared a sample of 353 barred Seyfert AGNs to 328 unbarred Seyferts and measure the potential correlation between the presence of the bar and the AGN. We find that at fixed mass and colour, AGN hosts show a small increase in the fraction of galaxies that are barred. The average difference is \db~$=0.012$, or roughly 16.0\% of the average barred AGN fraction. We find no difference in the $L_{\rm{[O~III]}}/$\mbh~ratio between barred and unbarred AGN at either fixed mass or colour. We conclude that while AGN hosts have moderately higher probabilities of hosting a bar, the presence of the bar does not seem to affect either the quantity or efficiency of fueling the central black hole.  

If bars are not required to initiate AGN fueling, then what is the source? There must be a process that transports angular momentum through the galactic disc and creates/maintains an accretion disc. Both theoretical models \citep{Shlos89,Shlos90} and numerical simulations \citep{HandQ10} indicate that this process requires two stages. First, the gas must be driven from a radial scale of megaparsecs down to kiloparsecs. Standard viscous torques on the gas are too inefficient to initiate gas inflow by themselves \citet{Shlos89,Bournaud05}; therefore, some other mechanism is required. Within the central kiloparsec, a secondary instability must take over within the gaseous disc for AGN fueling to occur. 

In the context of this general model, we consider three possibilities: (I) Bars are a necessary ingredient for fueling AGN, (II) Bars are one of multiple processes that fuel AGN, or (III) Bars play no role in fueling AGN. We also discuss in each scenario possible explanations for the existence of all four observed combinations: barred AGN, unbarred AGN, barred non-AGN, and unbarred non-AGN. 

\subsection{Scenario I: Bars are necessary to fuel AGN}\label{ssec:scenario1}

\noindent If the presence of a stellar bar is the only mechanism by which gas can be driven to the $\sim1$~kpc scale, there must be a reason both barred and unbarred AGN are observed in large numbers. One possibility is that a galactic bar \emph{initiates} fueling of the black hole, but is subsequently destroyed in a dynamic timescale shorter than the lifetime of the AGN. These separate timescales are not currently known with certainty, but estimates place the lifetime of an AGN from $10^6$ --- $10^8$ years \citep[eg,][]{Ski10,Martini04}. The range of bar lifetimes is not yet firmly established; some models show bars to be transient features that are destroyed either due to buckling from angular momentum transport or from the build-up of a central mass concentration (CMC) \citep{Bournaud05,Combes08}. In these models, the lifetime of a bar is estimated to be $1-2$~Gyr. \citet{Kraljic2012} also found bars to be short-lived in their simulations, but only early bars (formed at $z>1$). Bars formed later (at $z<1$) were maintained down to $z=0$, giving a lifetime of at least 8 Gyr. 

Other simulations \citep{Debattista04,Debattista06,Athanassoula05,Athanassoula13,Shen2004} do not observe bar destruction due to buckling. In these cases, only a sufficently massive CMC is capable of destroying the bar on Gyr timescales. The mass of the CMC required in these models is at least several percent of the total mass of the disc --- this is significantly larger than the mass measured in local disc galaxies. If the CMC is insufficently large, the bar is maintained for the lifetime of the disc \citep[up to 10~Gyr;][]{Athanassoula13}, and thus should be observable for at least the lifetime of the AGN. 

If bars are truly long-lived structures in all disc galaxies and are necessary to fuel AGN, we would expect a much higher value of the ratio of barred to unbarred AGN hosts. Since the observed numbers are nearly 1:1, we consider this scenario highly unlikely. It is possible that bars are necessary to fuel AGN, but the number of observed unbarred AGN can only be explained if the factor of $\sim10$ difference between the upper end of the AGN lifetime and the lower end of the bar lifetime can be resolved. While this is possible, we consider it unlikely given the assumptions required. 

\subsection{Scenario II: Bars are one of several ways to fuel AGN}\label{ssec:scenario2}

If stellar bars are only one of several ways to fuel AGN, then both barred and unbarred AGN should exist (as should both barred and unbarred star-forming galaxies). The simulations conducted by \citet{HandQ10} support this model, which show that multiple large-scale mechanisms (including a stellar bar) can be responsible for transporting gas to scales required for AGN fueling. Further, if bar-driven fueling is responsible for some fraction of the AGN, this model predicts an increase in the fraction of barred AGN, which our data supports.

While the existence of unbarred AGN is explained by this model, there is no immediate explanation for the existence of barred galaxies that do not host AGN; here we suggest several possibilities. First, a bar that initiates AGN fueling may simply outlive the AGN (see \ref{ssec:scenario1}), which agrees with estimates of both bar and AGN lifetimes. Second, there could be a correlation between bar strength and AGN activity, where only sufficiently strong bars initiate fueling. This is consistent with \citet{Lee12}, who find a higher AGN fraction in barred galaxies where the bar length is at least $1/4$ of the total disc diameter. They did not test, however, whether this relationship remains at fixed mass and colour. Finally, the emission from an AGN is expected to be highly variable with time, driven by processes such as accretion disc instabilities and/or feedback within the accreting material \citep{Hickox14}. In this case, barred galaxies without AGN are simply observed in low parts of their duty cycle, with Eddington ratios too low to be detected at the limits of our observations.

\subsection{Scenario III: Bars do not fuel AGN}\label{ssec:scenario3}

Finally, we consider the possibility that stellar bars do not trigger AGN activity in any way. This is inconsistent (although marginally so) with the increase in barred vs. unbarred AGN fractions that we find at fixed mass and colour. One possibility is that the model used for the null hypothesis (a normal distribution centered at \db$=0$) does not apply. Detailed simulations of cosmological volumes that include both AGN and detailed disc morphology, such as \textit{Illustris} \citep{Illustris} and EAGLE \citep{EAGLE} should ultimately provide more well-defined priors for this.

In addition, our test of the null hypothesis could still be consistent with a strong effect even if the total number of barred and unbarred bins were equal. For example, if bar-driven fueling is strongly mass-dependent, the \db~bins could have excesses of barred AGN at high masses and deficits at low masses; this would still be consistent with a distribution centered at zero. We test the simplest cases by simply splitting the sample into two in both mass and colour (Table~\ref{masscolortable}). Low- and high-mass disc galaxies (dividing the sample at $\log(M/M_\odot)=10.625$) have nearly identical values of \fb~and mean~\db; there is no evidence of a mass-dependent effect on bar-driven AGN fueling. When splitting discs into red vs. blue (at a colour of $(u-r)=2.22$), bluer galaxies do have significantly more bins with an excess of barred AGN (\fb$=0.88$) than redder galaxies (\fb$=0.54$). The uncertainties on \fb~are quite large, though, since each subsample has less than a dozen bins. Our splits by colour agree with \citep{Oh12}, who find that bar effects on AGN are more pronounced in bluer and less massive galaxies. \citet{Lee12}, in contrast, find that \fb~depends on neither mass nor colour.

\begin{table}
\begin{center}
\begin{tabular}{lccc}
\hline
\hline
            & sample                    & \fb     & Mean \db \\
\hline
low mass    &  $\log(M/M_\odot)<10.625$ &  0.70   & 0.0125   \\
high mass   &  $\log(M/M_\odot)>10.625$ &  0.64   & 0.0123   \\
\hline
blue         &  $(u-r) < 2.22$           &  0.88   & 0.023    \\
red        &  $(u-r) > 2.22$           &  0.54   & 0.006    \\
\hline
\hline
\end{tabular}
\caption{Difference between barred and unbarred AGN fractions for disc galaxies when splitting the sample in two by both mass and colour. \fb~is the fraction of bins that show an excess of barred AGN (compared to unbarred), while \db~is the average value of the differences over all bins. Since the number of bins in each subsample is only $\sim8-13$ when splitting by mass or colour, the uncertainty in \fb~is correspondingly large.}
\label{masscolortable}
\end{center}
\end{table}

If bars have no impact at all on the likelihood of a disc galaxy hosting an observable AGN, this is inconsistent with both the models and simulations that demonstrate efficient gas-driven inflow by bar structures \citep{HandQ10}. If the efficiencies of other morphologies that drive gas inflow are much higher than bars, though, this could also be consistent with our data. A lack of bar-driven fueling is consistent with the existence of both barred and unbarred AGN and star-forming galaxies, and the nearly equal numbers found in both pairs. 

Given the limits on the data set (which is driven by binning the total number of disc galaxies by mass and colour), we do not completely rule out this model. However, given the small (but measurable) increase in the bar fraction from our data and the current constraints on both bar and AGN timescales, we propose that bar-driven fueling must account for at least some fraction of observed AGN activity (\S\ref{ssec:scenario2}).

\section{Conclusions}
\label{sec:conclusions}
We have created a sample of 19,756 disc galaxies from SDSS DR7, using data from the \gztwo~project for morphological classifications of strong, large-scale bars. We studied the effects of stellar bars on 681~AGN and compared these effects to a control sample of disc galaxies both without bars and without AGN. The \gztwo~data provides a very large sample of disc morphologies for which the bar likelihood can be empirically quantified, based on crowdsourced visual classifications.

We find that the fraction of barred AGN (51\%) is significantly greater than the fraction of barred galaxies with central star formation (37\%). However, this is driven both by selection effects for detecting optically-identified AGN and by known correlations between black-hole mass and stellar mass, as well as stellar mass and optical colour. When examining the fraction of barred AGN as a function of a \underline{fixed} mass and colour, we still find a small increase in the number of barred AGN hosts. The null hypothesis of no relationship between the two cannot be ruled out at the 95\% confidence level. The $L_{\rm{[O~III]}}$/\mbh~ratio $R$ (a proxy for the overall accretion rate) shows no dependence on the presence of a bar, once the same mass and colour constraints are applied.

Our results are consistent with a small relationship between the presence of a large-scale galactic bar and the presence of an AGN. We propose that while bar-driven fueling does indeed contribute to some fraction of the current observed population of growing black holes, other dynamical mechanisms, such as lopsided or eccentric stellar disk, must also contribute to the redistribution of angular momentum and thus the fueling of the accretion disk at small galactic radii. 

Even with the advent of the large-scale SDSS data and the morphological classifications from \gztwo, this result is still constrained by the total number of galaxies in our study. Larger samples of disk galaxies with activity and morphological classifications, notably the Dark Energy Survey (DES) and the Large Synoptic Survey Telescope (LSST), should increase the sample sizes by factors of at least a few and help to confirm these results. Further development on the theoretical side is also critical --- with state-of-the-art simulations now able to reproduce both the morphology distributions and the observed black hole mass function, these results can be compared to theory in a cosmological context.


\section*{Acknowledgments}
The data in this paper are the result of the efforts of the Galaxy~Zoo~2 volunteers, without whom none of this work would be possible. Their efforts are individually acknowledged at \url{authors.galaxyzoo.org}. Please contact the author(s) to request access to research materials discussed in this paper. We thank Nathan Cloutier, Michael Rutkowski, and Claudia Scarlata for useful discussions. We also thank the referee for thoughtful comments which improved the quality of this paper.

The development of \gztwo~was supported by The Leverhulme Trust. MAG, KWW and LFF would like to acknowledge support from the US National Science Foundation under grant DRL-0941610 and from the UMN Grant-in-Aid program. KS gratefully acknowledges support from Swiss National Science Foundation Grant PP00P2\_138979/1. TM acknowledges funding from the STFC ST/J500665/1. RCN was partially supported by STFC grant ST/K00090X/1. BDS acknowledges support from Worcester College, Oxford, and from the Oxford Martin School program on computational cosmology.

This research made extensive use of the Tool for OPerations on Catalogues And Tables (TOPCAT), which can be found at \url{www.starlink.ac.uk/topcat/} \citep{tay05}. 

Funding for the SDSS and SDSS-II has been provided by the Alfred P. Sloan Foundation, the Participating Institutions, the National Science Foundation, the U.S. Department of Energy, the National Aeronautics and Space Administration, the Japanese Monbukagakusho, the Max Planck Society, and the Higher Education Funding Council for England. The SDSS website is \url{www.sdss.org}.

The SDSS is managed by the Astrophysical Research Consortium for the Participating Institutions. The Participating Institutions are the American Museum of Natural History, Astrophysical Institute Potsdam, University of Basel, University of Cambridge, Case Western Reserve University, University of Chicago, Drexel University, Fermilab, the Institute for Advanced Study, the Japan Participation Group, Johns Hopkins University, the Joint Institute for Nuclear Astrophysics, the Kavli Institute for Particle Astrophysics and Cosmology, the Korean Scientist Group, the Chinese Academy of Sciences (LAMOST), Los Alamos National Laboratory, the Max-Planck-Institute for Astronomy (MPIA), the Max-Planck-Institute for Astrophysics (MPA), New Mexico State University, Ohio State University, University of Pittsburgh, University of Portsmouth, Princeton University, the United States Naval Observatory, and the University of Washington.

\bibliographystyle{mn2e}
\bibliography{melrefs}  

\newpage
\clearpage
\vspace{0pc}
\begin{deluxetable}{lrrccccrcccccc}
\rotate
\tablewidth{0pc}
\tabletypesize{\scriptsize}
\tablecaption{GZ2 catalogue of ``not edge-on'' disc galaxies}
\tablehead{
 \colhead{SDSS DR7} & 
 \colhead{RA} &
 \colhead{Dec} &  
 \colhead{$M_{z,petro}$} &
 \colhead{$(u-r)$} &
 \colhead{Redshift} &  
 \colhead{BPT} &  
 \colhead{$M_{\rm star}$} &
 \colhead{$\sigma$}  &
 \colhead{$R$} &
 \colhead{$p_{\rm features}$} & 
 \colhead{\pnotedgeon} & 
 \colhead{\pbar} &
 \colhead{$N_{bar}$}
  \\ 
 \colhead{object ID} & 
 \colhead{J2000} &
 \colhead{J2000} &  
 \colhead{} &
 \colhead{} &
 \colhead{} &  
 \colhead{class} &  
 \colhead{$[\log(M_{\odot})]$} &
 \colhead{[km~s$^{-1}]$}  &
 \colhead{[$M_{\odot}/L_{\odot}$]} &
 \colhead{$_{\rm or~disk}$} & 
 \colhead{} & 
 \colhead{} &
 \colhead{}
}
\startdata
587742191520383064   &   196.5630 & 25.4605 &    $-21.26$    &  1.52   &  0.024  & 1    &   10.14   &  51.9  $\pm$ 5.4 & $0.9  $ &  0.839 & 1.000 & 0.689 & 27 \\
588010358534504619   &   146.0390 &  3.4041 &    $-19.72$    &  1.84   &  0.024  & 1    &   9.54    &   -              & -         &  0.918 & 0.763 & 0.433 & 29 \\
588017947743813765   &   192.2322 & 41.7789 &    $-21.22$    &  2.18   &  0.040  & 4    &   10.37   & 116.5  $\pm$ 5.8 & $-0.0 $ &  0.828 & 1.000 & 0.319 & 19 \\
587739828744945846   &   230.4270 & 20.5889 &    $-21.01$    &  2.59   &  0.042  & 3    &   10.45   &  90.8  $\pm$ 3.7 & $0.3  $ &  0.696 & 0.982 & 0.000 & 10 \\
587742062156382262   &   208.7955 & 21.3264 &    $-20.63$    &  1.57   &  0.029  & 1    &   9.99    &  57.5  $\pm$ 3.6 & $0.7  $ &  0.702 & 1.000 & 0.377 & 24 \\
587738570318675978   &   195.6816 & 15.5068 &    $-21.30$    &  1.62   &  0.022  & 1    &   10.27   &  65.0  $\pm$ 4.6 & $0.3  $ &  0.940 & 1.000 & 0.000 & 32 \\
588018055124746527   &   244.2345 & 32.9439 &    $-21.95$    &  3.12   &  0.031  & 0    &   10.85   & 194.0  $\pm$ 3.9 & $-1.3 $ &  0.931 & 0.893 & 0.978 & 34 \\
587726032763748389   &   130.4313 &  1.5305 &    $-21.78$    &  3.27   &  0.050  & 3    &   10.95   & 137.0  $\pm$ 5.7 & $-0.0 $ &  0.980 & 0.943 & 0.293 & 51 \\
587742550688596007   &   237.4456 & 12.3993 &    $-21.08$    &  2.08   &  0.015  & 3    &   10.22   &  83.0  $\pm$ 2.3 & $0.8  $ &  0.852 & 0.965 & 0.145 & 28 \\
587739098060619830   &   177.2420 & 37.5949 &    $-20.36$    &  1.38   &  0.038  & 1    &   9.81    &  46.5  $\pm$ 11.5 & $1.2 $ &  0.758 & 0.673 & 0.000 & 11 \\
587732470385279247   &   124.6074 & 29.9596 &    $-20.17$    &  1.99   &  0.020  & 1    &   9.89    &  58.1  $\pm$ 5.6 & $-0.7 $ &  0.583 & 1.000 & 0.293 & 21 \\
587742616172757184   &   241.3618 & 15.0253 &    $-21.58$    &  2.67   &  0.016  & 4    &   10.63   & 124.6  $\pm$ 1.9 & $-1.8 $ &  0.643 & 1.000 & 0.072 & 27 \\
588017978917257898   &   247.1737 & 21.5491 &    $-20.01$    &  1.94   &  0.038  & 1    &   9.79    &  -               & -         &  1.000 & 1.000 & 0.211 & 37 \\
587729157890900120   &   190.5697 &  4.0783 &    $-21.28$    &  2.10   &  0.048  & 2    &   10.45   &  98.6  $\pm$ 5.4 & $-0.0 $ &  0.650 & 0.923 & 0.000 & 13 \\
587739609173786650   &   210.1908 & 30.0760 &    $-22.01$    &  2.54   &  0.027  & 2    &   10.81   & 134.3  $\pm$ 3.3 & $-0.6 $ &  0.964 & 1.000 & 0.703 & 27 \\
588295842862334060   &   188.2374 & 50.9595 &    $-20.52$    &  1.78   &  0.041  & 1    &   9.90    &  42.2  $\pm$ 13.7& $0.6  $ &  0.980 & 1.980 & 0.672 & 38 \\
587742610805424255   &   255.7306 & 13.3583 &    $-21.48$    &  2.63   &  0.045  & 0    &   10.55   & 141.5  $\pm$ 3.3 & $-1.3 $ &  0.504 & 0.992 & 0.726 & 12 \\
587742550676668631   &   209.9203 & 18.0998 &    $-19.53$    &  2.52   &  0.037  & 1    &   9.87    &  25.3  $\pm$ 22.3& $1.5  $ &  0.963 & 0.987 & 0.585 & 51 \\
587739811035218261   &   240.5072 & 16.7577 &    $-19.80$    &  1.73   &  0.032  & 1    &   9.49    &   -              & -         &  0.815 & 1.000 & 0.047 & 26 \\
587733442659746092   &   240.6017 & 42.7571 &    $-20.24$    &  1.88   &  0.040  & 0    &   9.78    &  49.4  $\pm$ 22.5& $-0.2 $ &  0.926 & 0.965 & 0.211 & 33 \\
         \ldots \\
\enddata
\tablenotetext{1}{This table will be available in its entirety in machine-readable and Virtual Observatory (VO) forms in the online journal and on Vizier. A portion is shown here for guidance regarding its form and content. BPT classes are: 0 -- undetermined; 1 -- star-forming, 2 -- composite, 3 -- Seyfert AGN, 4 -- LINER.}
\label{tab:cat}
\end{deluxetable}

\end{document}